\documentclass[usenatbib]{an}
\usepackage{times,graphics,amssymb,epsfig,color}

\newcommand{\ber}{\begin{eqnarray}}
\newcommand{\eer}{\end{eqnarray}}

\def\apj{ApJ}

\def\labequn #1{\label{eq:#1}}
\def\labfig #1{\label{fig:#1}}
\def\labsecn #1{\label{sec:#1}}
\def\labsubsecn #1{\label{subsecn:#1}}
\def\labsubsubsecn #1{\label{subsubsecn:#1}}

\def\equn #1{Equation~\ref{eq:#1}}
\def\dequn#1#2{Equations~{\ref{eq:#1}}~and~{\ref{eq:#2}}}
\def\fig #1{Figure~\ref{fig:#1}}

\def\dsecn #1#2{Sections~{\ref{sec:#1}}~and~{\ref{sec:#2}}}
\def\subsecn #1{Section~\ref{subsecn:#1}}
\def\subsubsecn #1{Section~\ref{subsubsecn:#1}}

\def\etal{et al.\ }

\def\unit #1{\,{\rm #1}}

\def\kev{\unit{keV}}

\begin{document}
\title{Magneto centrifugal winds from accretion discs around black hole binaries}
%{Magnetically-driven winds in black hole X-ray binaries}
\author{S. Chakravorty\inst{1,2} \and P-O. Petrucci\inst{1,2} \and J. Ferreira\inst{1,2} \and G. Henri\inst{1,2} \and R. Belmont\inst{3,4} \and M. Clavel\inst{5} \and S. Corbel\inst{5} \and J. Rodriguez\inst{5} \and M. Coriat\inst{3,4} \and S. Drappeau\inst{3,4} \and J. Malzac\inst{3,4}}
\titlerunning{MHD winds in BHBs}
\authorrunning{Chakravorty \etal}
\institute{Univ. Grenoble Alpes, IPAG, F-38000 Grenoble, France\\e-mail: susmita.chakravorty@obs.ujf-grenoble.fr \and CNRS, IPAG, F-38000 Grenoble, France \and Universit\'e de Toulouse; UPS-OMP; IRAP, F-31028, Toulouse, France \and CNRS; IRAP; 9 Av. colonel Roche, F-31028, Toulouse, France \and Laboratoire AIM (CEA/IRFU - CNRS/INSU - Universit\'e Paris Diderot), CEA DSM/IRFU/SAp, F-91191 Gif-sur-Yvette, France}


%%%%%%%%%%%%%%%%%%%%%%%%%%%%%%%%%%%%%%%%%%%%%%%%%%%%%%%%

\abstract
{We want to test if self-similar magneto-hydrodynamic (MHD) accretion-ejection
models can explain the observational results for accretion disk winds in BHBs.
In our models, the density at the base of the outflow, from the accretion disk,
is not a free parameter, but is determined by solving the full set of dynamical
MHD equations without neglecting any physical term.  Different MHD solutions
were generated for different values of (a) the disk aspect ratio
($\varepsilon$) and (b) the ejection efficiency ($p$). We generated two kinds
of MHD solutions depending on the absence (cold solution) or presence (warm
solution) of heating at the disk surface. The cold MHD solutions are found to
be inadequate to account for winds due to their low ejection efficiency. The
warm solutions can have sufficiently high values of $p (\gtrsim 0.1)$ which is
required to explain the observed physical quantities in the wind. The heating
(required at the disk surface for the warm solutions) could be due to the
illumination which would be more efficient in the Soft state. We found that in
the Hard state a range of ionisation parameter is thermodynamically unstable,
which makes it impossible to have any wind at all, in the Hard state. Our
results would suggest that a thermo-magnetic process is required to explain
winds in BHBs.}

\maketitle

%%%%%%%%%%%%%%%%%%%%%%%%%%%%%%%%%%%%%%%%%%%%%%%%%%%%%%%%%%%%%%%%%%%

%%%%%%%%%%%%%%%%%%%%%%%%%%%%%%%%%%%%%%%%%%%%%%%%%%%%%%%%%%%%%%%%%%%%

%%%%%%%%%%%%%%%%%%%%%%%%%%%%%%%%%%%%%%%%%%%%%%%%%%%%%%%%%%%%%%%%%%%%
\section{Introduction}
\labsecn{sec:introduction}

High resolution X-ray spectra, from \textit{Chandra} and XMM-Newton, of stellar
mass black holes in binaries (BHBs) show blueshifted absorption lines. These
are signatures of winds from the accretion disk around the black hole (see
Neilsen and Homman, 2012 and references therein)). It has been, further, shown
for all BHBs that the absorption lines are more prominent in the Softer
(accretion disk dominated) states (Ponti \etal 2012 and references therein).

In this paper we investigate the magneto hydrodynamic
(hereafter MHD) solutions as driving mechanisms for winds from the accretion
disks around BHBs - cold solutions from Ferreira (1997, hereafter F97) and
warm solutions from Casse \& Ferreira (2000b) and Ferreira (2004).

%%%%%%%%%%%%%%%%%%%%%%%%%%%%%%%%%%%%%%%%%%%%%%%%%%%%%%%%%%%%%%%%%%%%

%%%%%%%%%%%%%%%%%%%%%%%%%%%%%%%%%%%%%%

\section{The MHD accretion disk wind solutions}
\labsecn{sec:MhdWinds}

We use the F97 solutions describing steady-state, axisymmetric solutions under
the following two conditions: (1) A large  scale  magnetic  field  of  bipolar
topology  is  assumed to thread the accretion disk. The strength of the
required vertical magnetic field component is obtained as a result of the
solution (Ferreira, 1995). (2) Some anomalous turbulent resistivity is at work,
allowing the plasma to diffuse through the field lines inside the disk.      
For a set of disk parameters, the solutions are computed from the disk midplane
to the asymptotic regime, the outflowing material becoming, first, super
slow-magnetosonic, then, Alfv\'enic and finally, fast-magnetosonic after which
they recollimate. In this paper we rely on those solutions only, which cross
their Alfv\'en surfaces before recollimating.

Because of ejection, the disk accretion rate varies with the radius even in a
steady state, namely $\dot M_{acc} \propto r^p$. This radial exponent, $p$
is very important
since it measures the local ejection efficiency. The larger the exponent, the
more massive and slower is the outflow. Mass conservation writes
\begin{equation}
n^+ m_p = \rho^+ \simeq \frac{p}{\varepsilon} \frac{\dot M_{acc}}{4 \pi \Omega_K r^3} 
\labequn{eqn:rho+}
\end{equation}
where $m_p$ is the proton mass and the superscript "+" stands for the height
where the flow velocity becomes sonic, $\Omega_K h = \varepsilon V_K$, where
$V_K=\Omega_K r=\sqrt{GM_{BH}/r}$ ($G$: gravitational constant) is the
keplerian speed and $\varepsilon= \frac{h}{r}$ is the disk aspect ratio, where
$h(r)$ is the vertical scale height at the cylindrical radius $r$. Thus, the
wind density is mostly dependent on $p$ and $\varepsilon$ for a given disk
accretion rate $\dot M_{acc}$.  

In the MHD models used in this paper the value of the exponent $p$ influences
the extent of magnetisation in the outflow which is defined as $\sigma^+ \simeq
\frac{1}{p} \left ( \frac{\Lambda}{1+ \Lambda} \right )$ (F97, Casse \&
Ferreira 2000a) where $\Lambda$ is the ratio of the torque due to the outflow
to the turbulent torque (usually referred to as the viscous torque). A
magnetically dominated self-confined outflow requires $\sigma^+ >1$.  The F97
outflow models have been obtained in the limit $\Lambda \rightarrow \infty$ so
that  the self-confined outflows carry away all the disk angular momentum and
thereby rotational energy with $\sigma^+ \simeq 1/p \gg 1$.

For the MHD outflow (with given $\varepsilon$ and $p$) emitted from the
accretion disk settled around a black hole, the important physical quantities
are given at any cylindrical (r,z) by 
\begin{equation}
n(r,z) = \frac{\dot m}{\sigma_T r_g} \left(\frac{r}{r_g}\right)^{(p-3/2)} f_n(y) 
%n(r) &\propto & \frac{\dot m}{\sigma_T r_g} \left(\frac{r}{r_g}\right)^{(p-3/2)} 
\labequn{eqn:n_scaling}
\end{equation}
\begin{equation}
v_i(r,z) = c \left(\frac{r}{r_g}\right)^{-1/2} f_{v_i}(y) \nonumber 
\end{equation}
\begin{equation}
B_i(r,z) = \left (\frac{\mu_o m_p c^2}{\sigma_T r_g}  \right )^{1/2}  \left(\frac{r}{r_g}\right)^{(-5/4 + p/2)} f_{B_i}(y) 
\labequn{eqn:B}
\end{equation}
\begin{equation}
\tau_{dyn}(r) = \frac{2\pi r_g}{c} \left(\frac{r}{r_g}\right)^{3/2} f_{\tau}(y) \nonumber
\end{equation}
where $\sigma_T$ is the Thomson cross section, $c$ the speed of light, $r_g = G
M_{BH} / c^2$ is the gravitational radius, $\mu_o$ the vacuum magnetic
permeability, $y = z/r$ the self-similar variable and the functions $f_X(y)$
are provided by the solution of the full set of MHD equations. In the above
expressions, $n$ is the proton number density and we consider it to be $\sim
n_H$ (the Hydrogen number density); $v_i$ (or $B_i$) is any component of the
velocity (or magnetic field) and $\tau_{dyn} = 1/div \bf{V}$ (where $\bf{V}$ is
the plasma velocity) is a measure of the dynamical time in the flow. The
normalized disk accretion rate used in \equn{eqn:n_scaling} is defined by
\begin{equation}
\dot m = \frac{\dot M_{acc}(r_g) \, c^2}{L_{Edd}} \nonumber
\labequn{eqn:mdot}
\end{equation}
where $L_{Edd}$ is the Eddington luminosity.
%\\{\color{red} Pop thinks it is $\dot M_{acc}(r_{in})$. I am not sure, because I believe Jon had put it as $r_g$. So I keep it as is and will change in a later version (post submission), if required} 

%%%%%%%%%%%%%%%%%%%%%%%%%%%%%%%%%%%%%%%%%%%%%%%%%%%%%%%%%%%%%%%%%%%%%%%%%%%%%%

%%%%%%%%%%%%%%%%%%%%%%%%%%%%%%%%%%%%%%%%%%%%%%%%%%%%%%%%%%%%%%%%%%%%%%%%%%%%%%
\section{Observational constrains}
\labsecn{sec:ObsConstrns}

%%%%%%%%%%%%%%%%%%%%%%%%%%%%%%%%%%%%%%
\subsection{The spectral energy distribution for the Soft and the Hard state}
\labsubsecn{subsec:SED}

%%%%%%%%%%%%%%%%%%%%%%%%%%%%%%%%%%%%%%%%%%%%%%%%%%%%%%%%%%%%%%%%%%%%%%%%%%%%%
\begin{figure}
\begin{center}
\includegraphics[scale = 1, width = 8 cm, trim = 0 125 0 0, clip, angle = 0]{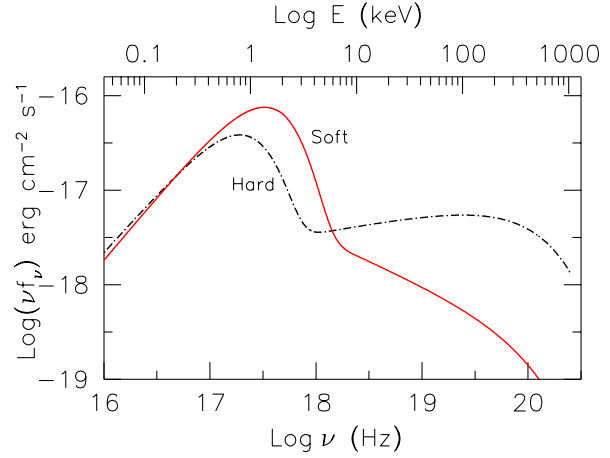}
\caption{The SEDs corresponding to the Soft and Hard states of the outburst of a
black hole of $10 M_{\odot}$. The two important components of the SED, namely,
the disk spectrum and the power-law have been added following the scheme
described in Remillard \& McClintock (2006).}
\labfig{fig:SED}
\end{center}
\end{figure}
%%%%%%%%%%%%%%%%%%%%%%%%%%%%%%%%%%%%%%%%%%%%%%%%%%%%%%%%%%%%%%%%%%%%%%%%%%%%%%

We follow the prescription given in Remillard \& McClintock (2006) to choose appropriate
values of the relevant parameters to derive the two representative SEDs for the
fiducial Soft and Hard states, for a black hole of $10 M_{\odot}$ for which
$r_g = 1.5 \times 10^6 \rm{cm}$.  
{\bf Soft state} (\fig{fig:SED} solid red curve): In the Soft
state the accretion disk extends all the way to $r_{in} = 3R_s = 6r_g$. Thus
$T(r_{in}) = 0.56 \kev$. The power-law has $\Gamma = 2.5$ and $A_{pl}$ is chosen
in such a way that the 2-20 keV disk flux contribution $f_d = 0.8$. 
{\bf Hard state} (\fig{fig:SED} dotted-and-dashed black
curve): With $r_{in} = 6R_s = 12r_g$ we generate a cooler disk with $T(r_{in}) =
0.33 \kev$. The power-law is dominant in this state with $\Gamma = 1.8$ and $f_d
= 0.2$.  
For each of the SEDs defined above, we use a high energy exponential cut-off so
that there is a break in the power-law at 100 keV.

For a $10 M_{\odot}$ black hole, the Eddington luminosity $L_{Edd}$ is $1.23
\times 10^{39} \rm{erg \, s^{-1}}$. We define the observational accretion rate
$\dot m_{obs} = L_{rad}/L_{Edd}$ where $L_{rad}$ is the 0.2 to 20 keV
luminosity.  $\dot m_{obs} = 0.14$ using the Soft SED and is equal to $0.07$
while using the Hard SED. Thus for simplicity we assume $\dot m_{obs} = 0.1$
for the rest of this paper. 

It is important to note here, the distinction between the disk accretion rate
$\dot{m}$ (\dequn{eqn:n_scaling}{eqn:mdot}) mentioned above, and the observed
accretion rate $\dot m_{obs}$ which is more commonly used in the literature.
One can define, 
\begin{equation}
\dot m = \frac{2}{\eta_{acc}} \frac{\dot m_{obs}}{\eta_{rad}}
\labequn{eqn:mdot_n_mdotobs}
\end{equation}
where the factor $2$ is due to the assumption that we see only one of the two
surfaces of the disk.  The accretion efficiency $\eta_{acc}\simeq r_g/2r_{in}$
depends mostly on the black hole spin. For the sake of simplicity, we choose
the Schwarzchild black hole, so that $\eta_{acc} \sim 1/12$, both in Soft and
Hard state.  The radiative efficiency, $\eta_{rad} = 1$ if the inner accretion
flow is radiatively efficient i.e. it radiates away all (or most) of the power
released due to accretion.  Thus $\dot m = 2.4$.

%%%%%%%%%%%%%%%%%%%%%%%%%%%%%

%%%%%%%%%%%%%%%%%%%%%%%%%%%%%%%%%%%%%%
\subsection{Finding the detectable wind within the MHD outflow}
\labsubsecn{subsec:DefWind}

%%%%%%%%%%%%%%%%%%%%%%%%%%%%%%%%%%%%%%%%%%%%%%%%%%%%%%%%%%%%%%%%%%%%%%%%%%%%%
\begin{figure}
\begin{center}
\includegraphics[scale = 1, width = 8 cm, trim = 0 240 0 0, clip, angle = 0]{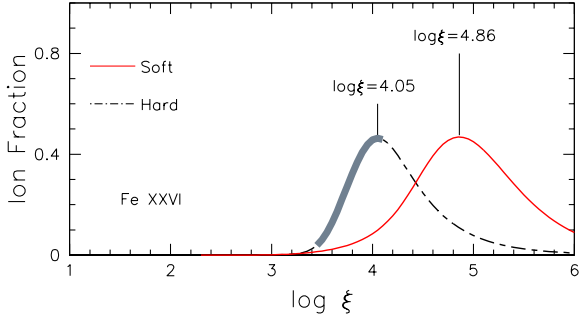}
\caption{The ion fraction distribution of FeXXVI with respect to $\log \xi$ is
shown for the two SEDs, Soft and Hard. The peak of the distribution is marked
and the corresponding $\log \xi$ values are labeled. Note that for the Hard
SED, a part of the distribution is highlighted by thick gray line - corresponds
to the thermodynamically unstable range of $\xi$. } 
\labfig{fig:If}
\end{center}
\end{figure}
%%%%%%%%%%%%%%%%%%%%%%%%%%%%%%%%%%%%%%%%%%%%%%%%%%%%%%%%%%%%%%%%%%%%%%%%%%%%%%

The MHD solutions can be used to predict the presence of outflowing material
over a wide range of distances. For any given solution, this outflowing material
spans large ranges in physical parameters like ionization parameter, density,
column density, velocity and timescales. Only part of this outflow will be
detectable through absorption lines - we refer to this part as the ``detectable
wind''.

%Ionization parameter 
We assume that at any given point within the flow, the gas is getting
illuminated by light from a central point source. As such, one of the key
physical parameters, in determining which region of the outflow can form a
wind, is the ionization parameter $\xi = L_{ion}/(n_H R^2_{sph})$
(Tarter et al. 1969). $L_{ion}$ is the luminosity of the ionizing light in the
energy range 1 - 1000 Rydberg (1 Rydberg = 13.6 eV) and $n_H$ is the density of
the gas located at a distance of $R_{sph}$

Detected ionized gas has to be thermodynamically stable. Photoionised gas in
thermal equilibrium will lie on the `stability' curve of $\log T$ vs
$\log(\xi/T)$ (Chakravorty et al. 2013, Higginbottom et al. 2015 and references
therein). If the gas is located (in the $\xi - T$ space) on a part of the curve
with negative slope then the system is considered thermodynamically unstable
because any perturbation (in temperature and pressure) would lead to runaway
heating or cooling. Thus we expect to detect gas which falls on the positive
slope part of the curve, because it will be thermodynamically stable and will
cause absorption lines in the spectrum. 

Using version C08.00 of CLOUDY\footnotemark (hereafter
C08, Ferland, 1998),
\footnotetext{URL: http://www.nublado.org/ }
we generated stability curves using both the Soft and the Hard SEDs as the
ionizing continuum. For the simulation of these curves we assumed the gas to
have solar metallicity, $n_H = 10^{10} \,\, \rm{cm^{-3}}$ and $N_H = 10^{23}
\,\, \rm{cm^{-2}}$. The Soft stability curve showed no unstable region, whereas
the Hard one had a distinct region of thermodynamic instability - $3.4 < \log
\xi < 4.1$. Thus, this range of ionization parameter has to be considered
undetectable, when we are using the Hard SED as the source of ionising light. 

%%%%%%%%%%%%%%%%%%%%%%%%%%%%%%%%%%%%%%%%%%%%%%%%%%%%%%%%%%%%%%%%%%%%%%%%%%%%%
\begin{figure}
\begin{center}
\includegraphics[scale = 1.0, width = 8 cm, trim = 0 0 0 0, angle = 0]{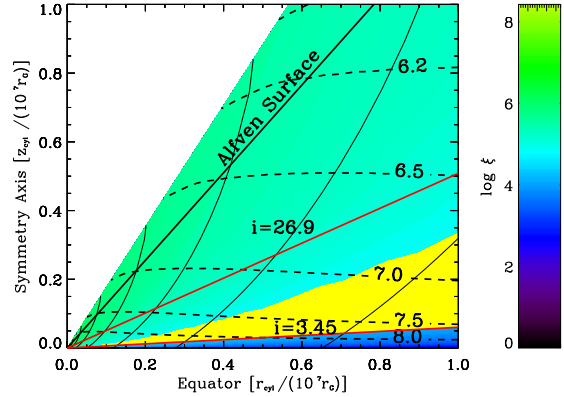}
\caption{\emph{Top Panel}: The distribution of the `Best Cold Set' in the plane
of the radial ($r_{cyl}$) and vertical ($z_{cyl}$) distance (in cylindrical
co-ordinates, and normalised to $10^7 r_g$) from the black hole. The colour
gradient informs about the $\xi$ distribution of the flow. The Alfv\'en surface
corresponding to the solution is also marked and labelled. The yellow wedge
highlights the wind part of the flow - this material is optically thin with
$N_H < 10^{24} \rm{cm^{-3}}$ and has sufficiently low ionization parameter
(with $\xi < 10^{4.86} \rm{erg \, cm}$) to cause FeXXVI absorption lines. The
angular extent of the wind is also clearly marked, where $i$ is the equatorial
angle. The dashed lines show the iso-contours of $n_H$, while the associated
labels denote the value of $\log n_H (\rm{cm^{-3}})$. }  
\labfig{fig:BestCold}
\end{center}
\end{figure}
%%%%%%%%%%%%%%%%%%%%%%%%%%%%%%%%%%%%%%%%%%%%%%%%%%%%%%%%%%%%%%%%%%%%%%%%%%%%%

We choose the presence of the ion FeXXVI as a proxy for detectable winds. The
probability of presence of the $X^{+i}$ ion is measured by its ion fraction
$I(X^{+i}) = \frac{N(X^{+i})}{f(X) \, N_{\rm{H}}}$, where $N(X^{+i})$ is the
column density of the $X^{+i}$ ion and $f(X) = n(X)/n_{\rm{H}}$ is the ratio of
the number density of the element $X$ to that of hydrogen.  \fig{fig:If} shows
that ion fraction of FeXXVI (calculated using C08) are, of course, different
based on whether the Soft or the Hard SED has been used as the source of
ionization for the absorbing gas.  The value of $\log \xi$, where the presence
of FeXXVI is maximised, changes from 4.05 for the Hard state, by $\sim$ 0.8
dex, to 4.86 for the Soft state. 

In the light of all the above mentioned observational constraints, we will
impose the following physical constraints on the MHD outflows (in
\dsecn{sec:EpPVar}{sec:WarmSolns}) to locate the detectable wind region within
them: \\ 
$\bullet$ In order to be defined as an outflow, the material needs to have positive
velocity along the vertical axis ($z_{cyl}$). \\ 
$\bullet$ Over-ionized gas cannot cause any absorption and hence cannot be
detected. Thus to be observable via FeXXVI absorption lines the ionization
parameter of the gas needs to have an upper limit. We imposed that $\xi \leq
10^{4.86} \,\, {\rm{erg \, cm}}$ (peak of FeXXVI ion fraction) for the Soft
state. For the Hard state, the constraint is $\xi \leq 10^{3.4} \,\, {\rm{erg
\, cm}}$, the value below which the thermal equilibrium curve is stable. \\
$\bullet$ The wind cannot be Compton thick and hence we impose that the integrated
column density along the line of sight satisfies $N_H < 10^{24}
{\rm{cm^{-2}}}$.

%%%%%%%%%%%%%%%%%%%%%%%%%%%%%%%%%%%%%%%%%%%%%%%%%%%%%%%%%%%%%%%%%%%%%%%%%%%%%
\begin{figure*}
\begin{center}
\includegraphics[scale = 1, height = 16 cm, trim = 85 10 125 20, clip, angle = 90]{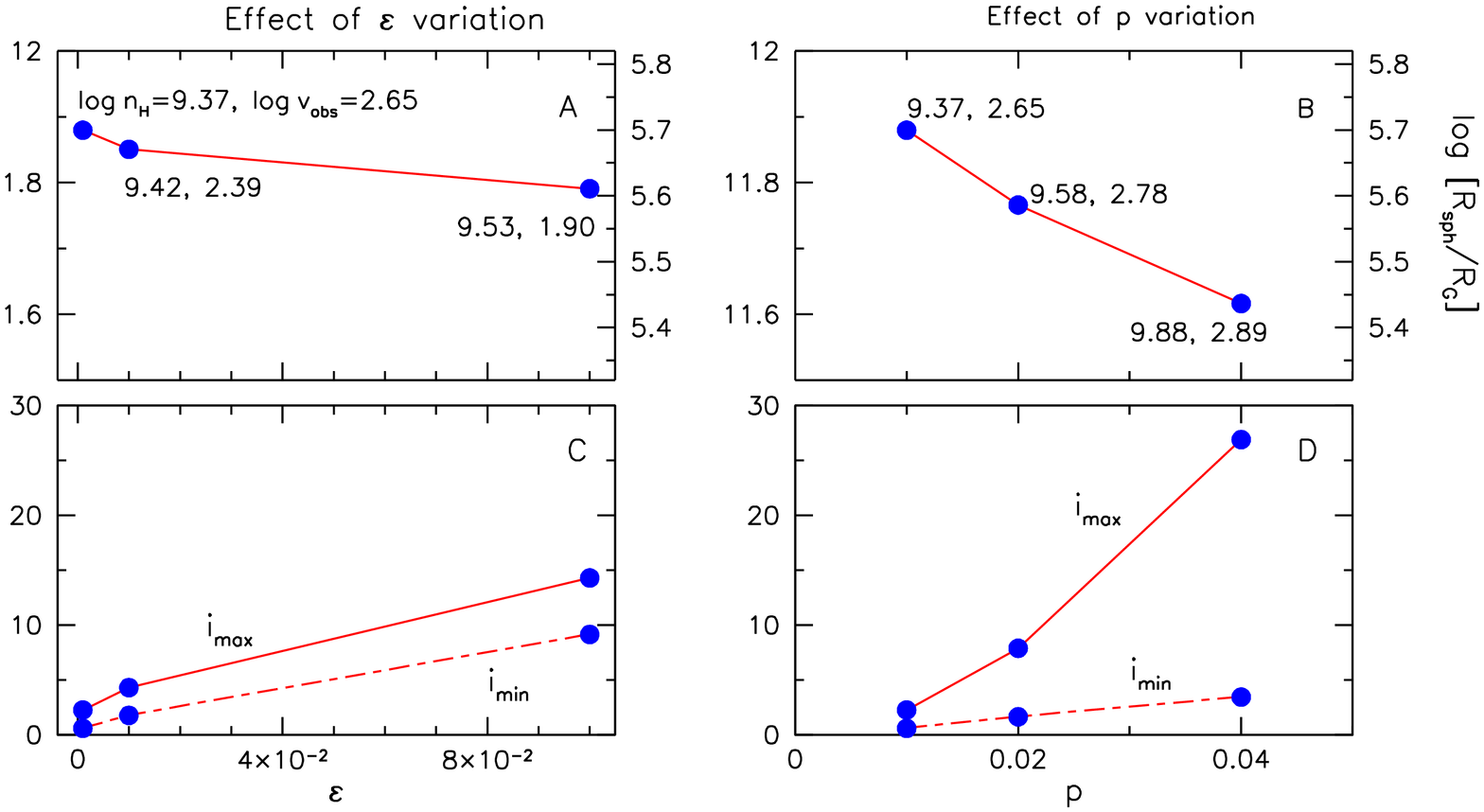}
\caption{The physical parameters of the wind are plotted as a function of
$\varepsilon$ (left panels) and $p$ (right panels), while using the Soft SED as
the ionizing continua. {\emph Top Panels}: For the closest wind point, we plot
the logarithm of $R_{sph}|_{wind}$ in the left panel A as a function of the
disk aspect ratio $\varepsilon$ and as a function of the accretion index $p$ in
the right panel B. $p = 0.01$ is held constant for the solutions in the left
panels and $\varepsilon = 0.001$ is kept constant for those in the right
panels. Each blue circle in the figure represents a MHD solution. The logarithm
of two other relevant quantities, $n_H$ and $v_{obs}$ for the closest wind
point are labeled at each point - these are their maximum possible values
within the wind region, for a given MHD solution. {\emph Bottom Panels}: The
minimum ($i_{min}$) and the maximum ($i_{max}$) equatorial angles of the line
of sight, within which the wind can be observed, is plotted as a function of
$\varepsilon$ (left) and of $p$ (right).} 
\labfig{fig:RsphDeltai}
\end{center}
\end{figure*}
%%%%%%%%%%%%%%%%%%%%%%%%%%%%%%%%%%%%%%%%%%%%%%%%%%%%%%%%%%%%%%%%%%%%%%%%%%%%%%

Here, we demonstrate how we choose the part of the MHD outflow
which will be detectable through absorption lines of FeXXVI.  For the
demonstration we use the MHD solution with $\varepsilon = 0.001$ and $p = 0.04$
which is illuminated by the Soft SED. Hereafter we will refer to this set of
parameters as the ``Best Cold Set''.  

We use the above mentioned physical constrains on the `Best Cold Set' and get
the yellow `wedge' region in \fig{fig:BestCold}. The wind is equatorial, for
the `Best Cold Set', not extending beyond $i = 26.9^{\circ}$.  The labelled
dashed black lines are the iso-contours for the number density $\log n_H
(\rm{cm^{-3}})$. We have checked that the velocities $v_{obs}$ within this
region fall in the range $10^2 - 10^3 \rm{km\, s^{-1}}$. We checked that
conditions of thermal equilibrium were satisfied within the wind region of the
outflow.     

This same method of finding the wind, and the associated physical conditions is
used for all the cold MHD solutions considered in this paper. In the subsequent
sections we will vary the MHD solutions (i.e. $\varepsilon$ and $p$) and
investigate the results using both the Soft and Hard SEDs. 

%%%%%%%%%%%%%%%%%%%%%%%%%%%%%%%%%%%%%%

%%%%%%%%%%%%%%%%%%%%%%%%%%%%%%%%%%%%%%%%%%%%%%%%%%%%%%%%%%%%%%%%%%%%%%%%%%%%%%
%\section{A phenomenological probe of the ``cold'' MHD solutions}
\section{The cold MHD solutions}
\labsecn{sec:EpPVar}

%%%%%%%%%%%%%%%%%%%%%%%%%%%%%%%%%%%%%%
\subsection{Effect of variation of the parameters of the MHD flow}
\labsubsecn{ParameterVar}

For observers, an important set of parameters are the distance
($R_{sph}|_{wind}$), density ($n_H|_{max}$) and velocity ($v_{obs}|_{max}$) of
the point of the wind closest to the black hole. Hereafter we shall call this
point as the `closest wind point'.  Note that for any given solution,
$n_H|_{max}$ and $v_{obs}|_{max}$ are the maximum attainable density and
velocity, respectively, within the wind region. Another quantity of interest
would be the predicted minimum $i_{min}$ and maximum $i_{max}$ equatorial
angles (of the line of sight) within which the wind can be observed. The
results are plotted in \fig{fig:RsphDeltai}.
 
$R_{sph}|_{wind}$ decreases, i.e. the winds goes closer to the black hole, as
$\varepsilon$ increases (panel A of \fig{fig:RsphDeltai}). $n_H|_{max}$
increases as $\varepsilon$ increases, but $v_{obs}|_{max}$ decreases (panel A).
The growth of $\Delta i = i_{max} - i_{min}$ with $\varepsilon$ (panel C) shows
that the wind gets broader as the disk aspect ratio increases. 

As $p$ increases, the wind moves closer to the black hole (panel B of
\fig{fig:RsphDeltai}). The total change in $n_H|_{max}$ is 0.51 dex as $p$
changes from 0.01 to 0.04 (as compared to 0.16 through $\varepsilon$
variation). Both $R_{sph}|_{wind}$ and $n_H|_{max}$ are effected more by the
variation in $p$ than by the variation in $\varepsilon$ The growth of $\Delta
i$ (panel D) is also higher as a function of increase in $p$, implying a higher
probability of detecting a wind when the flow corresponds to higher $p$ values.
Thus, $p$ is the relatively more dominant (compared to $\varepsilon$) disk
parameter to favour detectable winds.  

%%%%%%%%%%%%%%%%%%%%%%%%%%%%%%%%%%%%%%

%%%%%%%%%%%%%%%%%%%%%%%%%%%%%%%%%%%%%%
\subsection{Cold solutions for the Hard state} 
\labsubsubsecn{subsubsec:ColdHard}

For the entire range of $\varepsilon$ (0.001 - 0.1) and $p$ (0.1 - 0.4) we
analysed the MHD solutions illuminated by the Hard SED, as well. Note that for
the Hard SED, we have to modify the upper limit of $\xi$ according to the
atomic physics and thermodynamic instability considerations
(\subsecn{subsec:DefWind}). With the appropriate condition, $\log \xi \le 3.4$,
we could not find any wind regions within the Compton thin part of the outflow,
for any of the MHD solutions.  This is a very significant result, because this
provides strong support to the observations that BHBs do not have winds in the
Hard state.
 
%%%%%%%%%%%%%%%%%%%%%%%%%%%%%%%%%%%%%%

%%%%%%%%%%%%%%%%%%%%%%%%%%%%%%%%%%%%%%
\subsection{Cold solutions cannot explain observed winds} 
\labsubsubsecn{subsubsec:ColdFail}

For most of the observed BHB winds the reported density $\ge 10^{11}
\,\rm{cm^{-3}}$ the distance $\le 10^{10} \, \rm{cm}$ (Schulz \& Brandt, 2002;
Ueda et al. 2004; Kubota et al. 2007; Miller 2008; Kallman, 2009). Compared to
these observations, for even the `Best Cold Solution', $R_{sph}|_{wind}$ is too
high and $n_H|_{max}$ is too low. The same analysis indicates that a MHD
solutions with higher $\varepsilon$, say 0.01, and a high $p \ge 0.04$ would be
the better suited to produce detectable winds, comparable to observations.
However it is not possible to reach larger values of $p$ for the cold solutions
with isothermal magnetic surfaces.
 
Within the steady-state approach of near Keplerian accretion discs, the
magnetic field distribution is related to $p$ via \equn{eqn:B}. Note that these
Ferreira et al. MHD solutions, assume that the magnetic flux threading the disk
is a result of the balance between outward turbulent diffusion and inward
advection of the magnetic field. One might argue that cold solutions with
larger values of $p$ may be generated if the condition of the balance is
relaxed (i.e. if magnetic flux is either continuously advected inward, e.g. in
magnetized advection-dominated discs, or if magnetic flux continuously diffuses
outward). However, to relax the balance, one needs to relax either the
steady-state assumption or relax the Keplerian assumption. It is not clear that
whether self-similarity conditions will hold, if these aforementioned
assumptions are relaxed. Note that in the context of AGN, Fukumura et al.
(2010a, 2010b, 2014, 2015) have been able to reproduce the various components
of the absorbing gas using MHD outflows which would correspond to $p \simeq
0.5$, a value much higher than in our best cold solution. Looking at the MHD
solutions used by them (Contopoulos \& Lovelace, 1984), one cannot simply say
that it is the outcome of non-steady balance. Further, they have not relaxed
the steady-state assumption or the Keplerian assumption -  their solutions
remain self-similar out to $r_{out} = 10^6 r_{in}$. 

One way to get denser outflows with larger $p$, while keeping the assumption
which ensure self-similar solutions, is to consider some entropy generation at
the disc surface - this automatically leads to a magnetic field distribution
that is different from the usual Blandford \& Payne (1982) one. The disk
surface heating may be the result of illumination from the inner accretion disk
or of enhanced turbulent dissipation at the base of the corona. For such flows,
larger values of $p$ up to $\sim 0.45$ have been reported (Casse and Ferreira,
2000b; Ferreira, 2004).

%%%%%%%%%%%%%%%%%%%%%%%%%%%%%%%%%%%%%%

%%%%%%%%%%%%%%%%%%%%%%%%%%%%%%%%%%%%%%%%%%%%%%%%%%%%%%%%%%%%%%%%%%%%%%%%%%%%%

%%%%%%%%%%%%%%%%%%%%%%%%%%%%%%%%%%%%%%%%%%%%%%%%%%%%%%%%%%%%%%%%%%%%%%%%%%%%%%
\section{Warm MHD solutions}
\labsecn{sec:WarmSolns}

%%%%%%%%%%%%%%%%%%%%%%%%%%%%%%%%%%%%%%%%%%%%%%%%%%%%%%%%%%%%%%%%%%%%%%%%%%%%%
\begin{figure}
\begin{center}
\includegraphics[scale = 1.0, width = 8 cm, trim = 0 0 0 0, angle = 0]{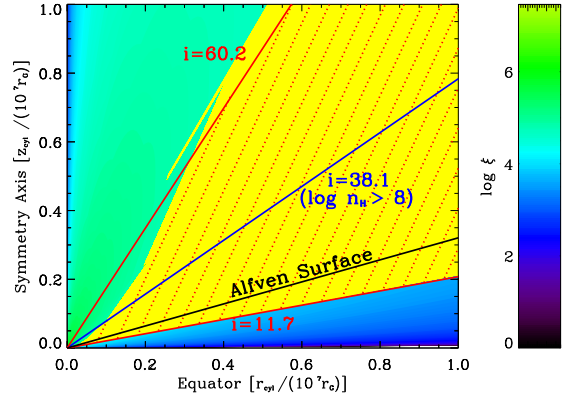}
\caption{The ionization parameter distribution for a Warm MHD solution with
$\varepsilon = 0.01$ and $p = 0.10$. The yellow region within the outflow is
obtained in the same way as in \fig{fig:BestCold}. The shaded region (with
dotted red lines) is the wind region within such a warm outflow - to obtain
this region we used the additional constraint that the cooling timescale of the
gas has to be lower than the dynamical time scale. Further, the solid blue line
with $i = 38.1^{\circ}$ is drawn to depict that high density material ($\log
n_H \ge 8.0$) in the flow is confined to low equatorial angles. }  
\labfig{fig:Warm}
\end{center}
\end{figure}
%%%%%%%%%%%%%%%%%%%%%%%%%%%%%%%%%%%%%%%%%%%%%%%%%%%%%%%%%%%%%%%%%%%%%%%%%%%%%

For the current analysis, we obtain dense warm solutions (with $p \ge 0.04$)
through the use of an ad-hoc heating function. We use the same shape for the
heating function, while playing only with its normalization - the larger the
heat input, the larger the value of $p$. For $\varepsilon = 0.01$ we could
achieve a maximum value of $p=0.11$.

\fig{fig:Warm} shows the wind for a Warm MHD solution with $p = 0.10$. The wind
(yellow region) spans a much wider range and extends far beyond the Alfv\'en
surface which was not the case for the cold MHD solutions. Hence we introduced
an additional constraint - the cooling timescale (calculated using C08) needs
to be shorter than the dynamical timescale - which was satisfied within the
yellow region if $i \le 60^{\circ}$. Thus the red-dotted shaded region is the
resultant detectable wind. However, note that the densest parts of the wind is
still confined to low equatorial angles - e.g. gas with $n_H \ge 10^8 \,
\rm{cm^{-3}}$ will lie below $i = 38.1^{\circ}$. 

We investigated warm MHD solutions with a range of $p$ values
(\fig{fig:WarmPhen}). $R_{sph}|_{wind}$ goes closer by a factor of 3.79 and
stands at $7.05 \times 10^4 \, r_g$, when $p$ increases from 0.04 to 0.11. The
highest density that we could achieve is $\log n_H = 11.1$ and the highest
velocity is $\log v_{obs} = 3.43$. Hereafter we shall refer to the $\varepsilon
= 0.01$ and $p = 0.10$ warm MHD solution as the ``Best Warm Solution''.

Clearly, warm solutions do a much better job than cold ones, as expected.
However, some observational results require the winds to have higher density
and lower distance than those produced by the ``Best Warm Solution''.  

We showed in \subsubsecn{subsubsec:ColdHard} that with the appropriate
restrictions (due to thermodynamic instability) on the $\xi$ value, no wind
could be found within the cold MHD outflows, in the Hard state. Since the warm
solutions result in much broader (than that in cold solutions) wind region, we
tested if the best warm solution can have a wind with a Hard SED. We use the
constraint that to be a detectable wind in the Hard state, the gas has to have
$\log \xi < 3.4$. Like the case of ``best cold solution'', here also we do not
find any wind region.  

%%%%%%%%%%%%%%%%%%%%%%%%%%%%%%%%%%%%%%

%%%%%%%%%%%%%%%%%%%%%%%%%%%%%%%%%%%%%%%%%%%%%%%%%%%%%%%%%%%%%%%%%%%%%%%%%%%%%%

%%%%%%%%%%%%%%%%%%%%%%%%%%%%%%%%%%%%%%%%%%%%%%%%%%%%%%%%%%%%%%%%%%%%%%%%%%%%%%
\section{Discussions and Conclusions}
\labsecn{sec:discussion}

%%%%%%%%%%%%%%%%%%%%%%%%%%%%%%%%%%%%%%
\subsection{Choice of upper limit of $\xi$}
\labsubsecn{subsec:UpperXi}

We used the limit $\log \xi \le 4.86$ to define the detectable wind. Note that
for the Soft SED, $\log \xi = 4.86$ corresponds to the peak of the ion fraction
of FeXXVI (\fig{fig:If}). The ion can have significant presence at higher
$\xi$. 

For the best warm solution we calculated the physical parameters for the
closest wind point for $\log \xi \le 6.0$. We find that $R_{sph}|_{wind}$
decreases by a factor of 93.4 bringing this point to $9.1 \times 10^2 r_g$. The
density at this point is $\log n_H = 13.71$ and the velocity is $\log v_{obs} =
4.28$. Thus we see that the parameters of closest point is sensitively
dependant on the choice of the upper limit of $\xi$. 

%%%%%%%%%%%%%%%%%%%%%%%%%%%%%%%%%%%%%%

%%%%%%%%%%%%%%%%%%%%%%%%%%%%%%%%%%%%%%%%%%%%%%%%%%%%%%%%%%%%%%%%%%%%%%%%%%%%%
\begin{figure}
\begin{center}
\includegraphics[scale = 1.0, width = 8 cm, trim = 0 260 0 0, angle = 0]{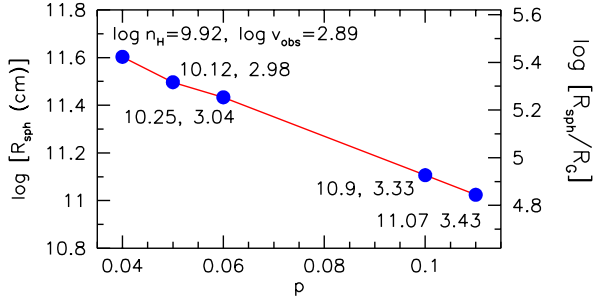}
\caption{Distance (density and velocity) of the closest wind point is (are)
plotted (labelled) as a function of $p$ for all the warm MHD solutions that we
investigated. $\varepsilon = 0.01$ is constant.}  
\labfig{fig:WarmPhen}
\end{center}
\end{figure}
%%%%%%%%%%%%%%%%%%%%%%%%%%%%%%%%%%%%%%%%%%%%%%%%%%%%%%%%%%%%%%%%%%%%%%%%%%%%%

%%%%%%%%%%%%%%%%%%%%%%%%%%%%%%%%%%%%%%
\subsection{The need for denser warm solution}
\labsubsecn{subsec:DenserWarmSoln}

As mentioned before, the Fukumura et.al. papers use MHD solutions with $p \simeq 0.5$ to model AGN outflows. We have not been able to
reproduce such high values of $p$ and are limited to $p = 0.11$, at present.
Our calculations show that as $p$ increased from 0.04
to 0.11 for the warm MHD solution, $R_{sph}|_{wind}$ for the closest wind point
decreased by a factor of 3.79.  Thus a further increase to $p \simeq 0.5$ may
take the closest wind point nearer to the black hole (hypothetically) by a
further factor of $\sim 10$, to $\sim 5 \times 10^3 r_g$ (assuming an almost
linear change in density as $p$ increases). We shall report the exact
calculations in our future publications. 

%%%%%%%%%%%%%%%%%%%%%%%%%%%%%%%%%%%%%%

%%%%%%%%%%%%%%%%%%%%%%%%%%%%%%%%%%%%%%%%%%%%%%%%%%%%%%%%%%%%%%%%%%%%%%%%%%%%%

%%%%%%%%%%%%%%%%%%%%%%%%%%%%%%%%%%%%%%%%%%%%%%%%%%%%%%%%%%%%%%%%%%%%%%%%%%%%%
\section{Conclusions}
\labsecn{conclusions}

In this paper we investigated if magneto centrifugal outflows (Ferreira, 1997;
Casse \& Ferreira 2000b) can reproduce the observed winds. The investigations
are done as a function of the two key accretion disk parameters - the disk
aspect ratio $\varepsilon$ and the radial exponent $p$ of the accretion rate
($\dot{M}_{acc} \propto r^p$). The results are summarised below: \\
$\bullet$ We need high values of $p (> 0.04)$ to
reproduce winds that can match observations. However $p$ cannot be increased to
desirable values in the framework of the cold MHD solutions. We definitely need
warm MHD solutions to explain the observational results. \\
$\bullet$ In the Soft state, our densest warm MHD solution predicts a
wind at $7.05 \times 10^4 r_g$ with a density of $\log n_H = 11.1$. The densest part
of the wind ($\log n_H > 8$) still remains equatorial - within $i \sim
30^{\circ}$ of the accretion disk. The values of the physical parameters are
consistent with some of the observed winds in BHBs. \\
$\bullet$ The outflow illuminated by a Hard SED will not produce
detectable wind because the wind region falls within the thermodynamically
unstable range of $\log \xi$ and hence unlikely to be detected. Further in the
absence of favourable illumination, it is likely that the Hard state will have
an associated cold outflow, which is incapable of producing the usually
observed winds. When these two aspects are considered together, we realise that
it is impossible to ever produce a wind in the canonical Hard state.

%%%%%%%%%%%%%%%%%%%%%%%%%%%%%%%%%%%%%%%%%%%%%%%%%%%%%%%%%%%%%%%%%%%%%%%%%%%%%%

%%%%%%%%%%%%%%%%%%%%%%%%%%%%%%%%%%%%%%%%%%%%%%%%%%%%%%%%%%%%%%%%%%%%%%%%%%%%%%
\acknowledgements{
The authors acknowledge funding support from the French Research National
Agency (CHAOS project ANR-12-BS05-0009 http://www.chaos-project.fr) and CNES.
This work has been partially supported by a grant from Labex OSUG@2020
(Investissements d'avenir – ANR10 LABX56)}
%\end{acknowledgements}

%%%%%%%%%%%%%%%%%%%%%%%%%%%%%%%%%%%%%%%%%%%%%%%%%%%%%%%%%%%%%%%%%%%%%%%%%%%%%%

%%%%%%%%%%%%%%%%%%%%%%%%%%%%%%%%%%%%%%%%%%%%%%%%%%%%%%%%%%%%%%%%%%%%%%%%%%%%%%
%\bibliographystyle{aa}

%%%%%%%%%%%%%%%%%%%%%%%%%%%%%%%%%%%%%%%%%%%%%%%%%%%%%%%%%%%%%%%%%%%%%%%%%%%%%%

\end{document}